\documentclass[12pt]{article}

\usepackage{graphics,graphpap}

\newcommand{\bea}{\begin{eqnarray}}
\newcommand{\eea}{\end{eqnarray}}
\newcommand{\beq}{\begin{equation}}
\newcommand{\eeq}{\end{equation}}

\def\msbar{\ifmmode{\overline{\rm MS}} \else{$\overline{\rm MS}$} \fi}
\def\drbar{\ifmmode{\overline{\rm DR}} \else{$\overline{\rm DR}$} \fi}
\def\sf{\ifmmode{\tilde{f}} \else{$\tilde{f}$} \fi}
\def\st{\ifmmode{\tilde{t}} \else{$\tilde{t}$} \fi}
\def\sb{\ifmmode{\tilde{b}} \else{$\tilde{b}$} \fi}
\def\sq{\ifmmode{\tilde{q}} \else{$\tilde{q}$} \fi}
\def\sg{\ifmmode{\tilde{g}} \else{$\tilde{g}$} \fi}
\def\bbar{\ifmmode{\bar{b}} \else{$\bar{b}$} \fi}
\def\tbar{\ifmmode{\bar{t}} \else{$\bar{t}$} \fi}
\def\qbar{\ifmmode{\bar{q}} \else{$\bar{q}$} \fi}

\newcommand\bvec{\left( \begin{array}{c}}
\newcommand\evec{\end{array}\right)}

\newcommand\bmat{\left( \begin{array}{cc}}
\newcommand\emat{\end{array}\right)}

\newcommand\ch{{\tilde{\chi}}}
\renewcommand\d{\delta}

\def\psla{{p \hspace{-1.8mm} \slash}}

\newcommand\ca{{\cos{\alpha}}}
\newcommand\sa{{\sin{\alpha}}}

\newcommand\cbe{{\cos{\beta}}}
\newcommand\sbe{{\sin{\beta}}}

\renewcommand\d{\delta}

\def\su{\ifmmode{\tilde{u}} \else{$\tilde{u}$} \fi}
\def\sd{\ifmmode{\tilde{d}} \else{$\tilde{d}$} \fi}

\newcommand{\DRbar}{{\overline{\rm  DR}}}

\newcommand{\Gtree}{\Gamma^{\rm tree}}
\newcommand{\Gnaive}{\Gamma^{\rm naive\,tree}}
\newcommand{\Gcorr}{\Gamma^{\rm corr.}}

\def\MSbar{{\overline{\rm MS}}}
\def\DRbar{{\overline{\rm DR}}}

\begin{document}
%------------------------------------------------------------------------

\pagestyle{empty} \vspace*{-1cm}
\begin{flushright}
  HEPHY-PUB 788/04 \\
  TU-716 \\
  hep-ph/0405187 \\
% Version 18052004
\end{flushright}

\vspace*{1.4cm}

\begin{center}
\begin{Large} \bf
Full one-loop corrections to %\\[2mm]
 SUSY Higgs boson decays into charginos
\end{Large}

\vspace{10mm}

{\large H. Eberl$^a$, W. Majerotto$^a$,
 Y.~Yamada$^b$}

\vspace{6mm}
\begin{tabular}{l}
 $^a${\it Institut f\"ur Hochenergiephysik der \"Osterreichischen
 Akademie der Wissenschaften,}\\
 \hphantom{$^a$}{\it A-1050 Vienna, Austria}\\
 $^b${\it Department of Physics, Tohoku University,
Sendai 980-8578, Japan}
\end{tabular}

\vspace{20mm}

\begin{abstract}
We present the decay widths of the heavier Higgs bosons ($H^0$, $A^0$) 
into chargino pairs 
in the minimal supersymmetric standard model, 
including full one-loop corrections. 
All parameters for charginos are renormalized 
in the on-shell scheme. The importance of the corrections to 
the chargino mass matrix and mixing matrices is pointed out. 
The full corrections are typically of the order of 10 \%. 
\end{abstract}
\end{center}

\vfill

\newpage
\pagestyle{plain} \setcounter{page}{2}

\section{Introduction}
The Minimal Supersymmetric Standard Model (MSSM) \cite{mssm} is
considered the most attractive extension of the Standard Model. This 
model contains two Higgs scalar doublets, 
implying the existence of five physical Higgs bosons \cite{gunion}; 
two CP-even neutral bosons ($h^0$, $H^0$), one CP-odd boson $A^0$, 
and two charged bosons $H^{\pm}$. For the verification of the MSSM, 
detection and precision studies of these Higgs bosons are 
necessary. 

The decay modes of the heavier Higgs bosons ($H^0$, $A^0$) are 
in general complicated \cite{tree1,tree2}, 
especially if $\tan\beta$, 
the ratio of the vacuum expectation values of the two Higgs scalars, 
is not much larger than one. 
For example, they may decay into pairs of the SUSY 
particles \cite{tree1} such as squarks, sleptons, charginos, 
and neutralinos. In this paper, we focus our attention on the 
decays into charginos, 
\begin{equation}
  (H^0, A^0) \to \ch^+_i  + \ch^-_j \, ,
  \label{eq:Hk0cha}
\end{equation}
with $i,j=(1,2)$. 
Existing numerical analyses \cite{tree1,tree2,tree3} at tree-level 
have shown that the decays (\ref{eq:Hk0cha}) have in general 
non negligible branching ratios. 
These decays are also interesting because 
they are generated by gaugino-higgsino-Higgs boson couplings~\cite{gunion} 
at tree-level and very sensitive to the components of charginos. 
Detailed studies of these decays would therefore 
provide useful information about the chargino sector, 
complementary to the pair production processes
$e^+e^-\hspace{-3pt}\rightarrow\ch^+_i\ch^-_j$~\cite{chaproduction}.

Since the masses and mixing matrices of the charginos are expected to 
be precisely determined at future 
colliders \cite{charginomeasure,LCs,LCexp}, 
it is interesting to study the radiative corrections to the 
decays~(\ref{eq:Hk0cha}). 
The one-loop corrections involving quarks and squarks in 
the third generation were calculated in Ref.~\cite{Zhang}. 
However, for the masses and mixings of the charginos, 
the corrections from quark-squark loops \cite{chmasscorr1} and 
those from the other loops \cite{chmasscorr3,chmasscorr2} are 
shown to be numerically comparable. It is therefore necessary 
to include the other loop corrections to the decays~(\ref{eq:Hk0cha}). 

In this paper, we study the widths of the decays~(\ref{eq:Hk0cha}) 
including full one-loop corrections and present numerical 
results for the $i=j=1$ case. We adopt the on-shell renormalization 
scheme for the chargino sector, following 
Refs.~\cite{chmasscorr1,chmasscorr2}. 
We also show numerical results for the one-loop corrected widths of 
the crossed-channel decay 
\begin{equation}
 \ch^{\pm}_2 \to \ch^{\pm}_1 + h^0 \, , 
 \label{eq:chidecay}
\end{equation}
which has been studied at tree-level \cite{inodecay}.

\section{Tree-level widths}

The tree-level widths for the decay $H_k^0 \to \ch^+_i \ch^-_j$, 
with $H^0_{\{1,2,3\}}\equiv\{h^0, H^0, A^0\}$ 
and $i,j=(1,2)$, are given by \cite{tree1}
\newpage
\begin{eqnarray}
&& \hspace{-5mm} \Gamma^{\rm tree}(H_k^0 \to \ch^+_i \ch^-_j) =
\frac{g^2}{16 \pi\, m^{3}_{H_{k}^0} }\,
\kappa(m_{H_{k}^0}^2,m^{2}_i,m^{2}_j) \,
\nonumber \\ 
&& \hspace{5mm}  \times
\left[ 
\left( m^{2}_{H_{k}^0} - m^{2}_i - m^{2}_j \right) 
(F_{ijk}^2 + F_{jik}^2 ) \, 
 - 4 \eta_k m_i m_j F_{ijk} F_{jik}  \right] \, ,
\label{eq:gammatree}
\end{eqnarray}
with $\kappa(x,y,z)\equiv ((x-y-z)^2-4yz)^{1/2}$. 
$\eta_k$ represents the CP eigenvalue of $H^0_k$; $\eta_{1,2}=1$ for the 
$(h^0,H^0)$ decays and $\eta_3=-1$ for the $A^0$ decays. 
We use the abbreviation $m_i\equiv m_{\ch^{\pm}_i}$. 
In this paper, we assume that the contributions of CP violation and 
generation mixings of the quarks and squarks are negligible. 

The chargino-Higgs boson couplings $gF_{ijk}$, defined 
by the interaction lagrangian 
\begin{equation}
  {\cal L}_{\rm int} = 
 -g\, H^0_a\,\overline{\ch^+_i} (F_{ija} P_R + F_{jia} P_L) \ch^+_j 
  +ig\, H^0_c\,\overline{\ch^+_i} (F_{ijc} P_R - F_{jic} P_L) \ch^+_j \, ,
\label{eq:Fchlag}
\end{equation}
with $a = 1,2$, $c = 3,4$, are given by \cite{gunion} 
\begin{eqnarray}
 gF_{ijk}&=& \frac{g}{\sqrt{2}} 
\left(  e_k\, V_{i1}U_{j2} - d_k\, V_{i2}U_{j1} \right) \,.
 \label{eq:Fchtree}
\end{eqnarray}
The would-be Nambu-Goldstone boson $H_4^0\equiv G^0$ is included 
here for later convenience. 
The mixing matrices ($U$, $V$) for the charginos are determined by 
diagonalizing the chargino mass matrix $X$ as 
\begin{equation}
 X = \left( \begin{array}{cc} M & \sqrt{2}\, m_W \sin\beta \\
                 \sqrt{2}\, m_W \cos\beta & \mu \end{array} \right) 
= U^{\dagger} \left( \begin{array}{cc} m_{\ch^+_1} & 0 \\
                 0 & m_{\ch^+_2} \end{array} \right) V 
\, .
\label{eq:chi+mat}
\end{equation}
Here $M$ and $\mu$ are the mass parameters of the SU(2) gaugino and 
higgsino states, respectively. We choose $U$ and $V$ to be real. 
The effect of the mixings of $H_k^0$ is represented by $e_k$ and $d_k$, 
which take the values 
 \begin{eqnarray}
 e_k&=&
 \Big(-\sa,\,\hphantom{-}\ca, \,-\sbe,\,\hphantom{-}\cbe\Big)_k\, , 
 \nonumber\\
 d_k&=&
 \Big(-\ca,\,-\sa,\,\hphantom{-}\cbe,\,\hphantom{-}\sbe\Big)_k\, . 
 \label{eq:dudd}
 \end{eqnarray}

We also show the widths of the decays $\ch^+_2\to\chi^+_1H_k^0$ 
at the tree-level \cite{inodecay}
\begin{eqnarray}
&& \hspace{-5mm} \Gamma^{\rm tree}(\ch^+_2\to\chi^+_1H_k^0 ) =
\frac{g^2}{32 \pi\, m^{3}_{\ch^+_2} }\,
\kappa(m^{2}_2,m^{2}_1,m_{H_{k}^0}^2) \,
\nonumber \\ 
&& \hspace{5mm}  \times
\left[ 
\left( m_2^2 + m_1^2 - m^{2}_{H_{k}^0} \right) 
(F_{12k}^2 + F_{21k}^2 ) \, 
 + 4 \eta_k m_1 m_2 F_{12k} F_{21k}  \right] \, .
\end{eqnarray}

\section{One-loop corrections}
We calculate the full one-loop corrections to the decay 
widths (\ref{eq:gammatree}). 

The one-loop correction to the coupling $F_{ijk}$ is expressed as 
\begin{equation}
  \label{eq:Fren}
F^{\rm corr.}_{ijk} = F_{ijk} + \Delta F_{ijk}
= F_{ijk}+ \d F^{(v)}_{ijk} + \d F^{(w)}_{ijk} +
 \d F^{(c)}_{ijk} \, , 
\end{equation}
where $\d F^{(v)}_{ijk}$, $\d F^{(w)}_{ijk}$, and $\d F^{(c)}_{ijk}$ 
are the vertex correction, the wave function correction, and the 
counter terms for the parameters in Eq.~(\ref{eq:Fchtree}), respectively. 

\begin{figure}
 \begin{center}
 \mbox{\resizebox{14cm}{!}{\includegraphics{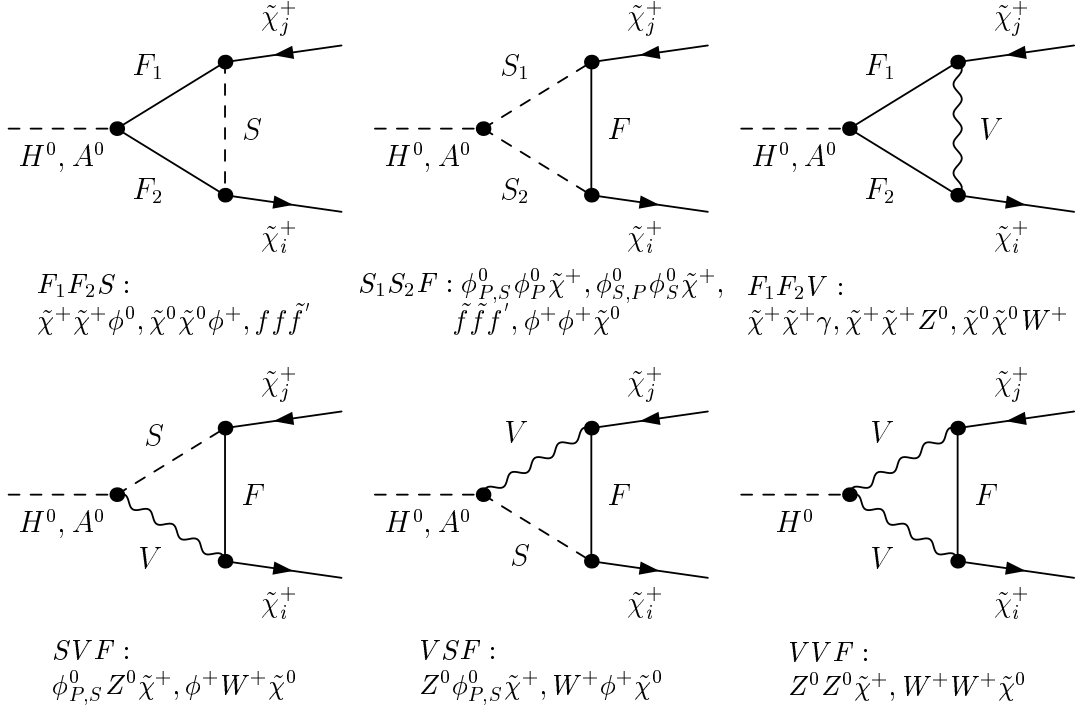}}}
  \end{center}
  \caption[feynman]{One-loop vertex corrections 
  to the $H^0_k\to\ch^+_i\ch^-_j$ decays, 
  $\phi^0 = \{\phi^0_S, \phi^0_P\} = \{h^0, H^0, A^0, G^0\}$,
  $\phi^+ = \{H^+, G^+\}$.
  \label{fig:feynman}}
\end{figure}

The vertex correction $\d F^{(v)}_{ijk}$ comes from the diagrams 
listed in Fig.~\ref{fig:feynman}. 
In this paper we do not show the analytic forms of these diagrams.  

The wave-function correction $\d F^{(w)}_{ijk}$ is expressed as 
\begin{equation}
 \delta{F}_{ijk}^{(w)}\,=\, \frac{\,1}{\,2}\,\bigg[\,
\delta{Z}^{H^0}_{lk}{F}_{ijl}+ \delta{Z}_{i'i}^{+L}{F}_{i'jk}
+\delta{Z}_{j'j}^{+R}{F}_{ij'k}\bigg]
 \,,
 \label{eq:dFw}
\end{equation}
with the implicit summations over $l = 1,2$ for k = 1 or 2, $l =
3,4$ for $k = 3$, and $i',j'= (1,2)$. The correction terms 
$\delta{Z}^{+(L,R)}$ for the chargino wave-functions are given 
by 
\begin{eqnarray}
%\hspace*{-1cm}  
\lefteqn{\delta{Z}^{+L}_{ii}  = } && \nonumber\\
&& \hspace*{-5mm} - {\rm Re}\,
 \bigg\{\Pi^{\ch L}_{ii}(m_i^2)+ m_i\,
 \Big[m_i\dot{\Pi}^{\ch L}_{ii}(m_i^2)+m_i\dot{\Pi}^{\ch R}_{ii}(m_i^2)
 +2\dot{\Pi}^{\ch S,L}_{ii}(m_i^2) 
 \Big]\bigg\}\, ,
 \label{eq:dZchpp}
 \\
 %[3mm]
%\hspace*{-1cm}  
\lefteqn{  \delta{Z}^{+L}_{pi} = } && \nonumber \\
&& \hspace*{-5mm}\frac{2}{m_p^2-m_i^2}\;
 {\rm Re}\left\{
 m_i^2 \Pi^{\ch L}_{pi}(m_i^2) + m_i m_p \Pi^{\ch R}_{pi}(m_i^2) + 
 m_p \Pi^{\ch\,S,L}_{pi}(m_i^2) + m_i \Pi^{\ch\,S,R}_{pi}(m_i^2) \right\} 
\, ,
 \label{eq:dZchps}
\end{eqnarray}
where $p\neq i$ and 
\begin{equation}
\Pi^{\ch}_{ij}(p)=\Pi^{\ch L}_{ij}(p^2)\psla P_L 
+\Pi^{\ch R}_{ij}(p^2)\psla P_R 
+\Pi^{\ch\,S,L}_{ij}(p^2)P_L+\Pi^{\ch\,S,R}_{ij}(p^2)P_R \, ,
\label{eq:charginoselfe} 
\end{equation}
are the self-energies of the charginos. 
$\delta Z^{+R}$ are obtained from Eqs.~(\ref{eq:dZchpp}, \ref{eq:dZchps})
by the exchange $L\leftrightarrow R$. 
The CP symmetry relation ${\rm Re}\Pi^{\ch S,L}_{ii}={\rm Re}\Pi^{\ch S,R}_{ii}$ 
is used in Eq.~(\ref{eq:dZchpp}). 
The corrections $\delta{Z}^{H^0}$ for the Higgs bosons are
\begin{eqnarray}
 \label{eq:dZHkk}
 \delta{Z}^{H^0}_{kk}&=&
  -\;{\rm Re}\,\dot{\Pi}^{H^0}_{kk}(m_{H_k^0}^2)\, , 
\hspace{30mm} k=1,2,3, \\ 
\label{eq:dZHlk}
\delta{Z}^{H^0}_{ab}&=& \frac{2}{m^2_{H^0_a}-m^2_{H^0_b}}\, {\rm
Re}\, \Pi^{H^0}_{ab}(m^2_{H^0_b}) \, , 
\hspace{10mm} a,b=(1,2), \;  a\neq b \\
\label{eq:dZAG}
\delta{Z}^{H^0}_{43}&=& -\frac{2}{m_{A^0}^2}\, {\rm
Re}\, \Pi^{H^0}_{43}(m_{A^0}^2) \, . 
\end{eqnarray}
The Higgs boson self-energies $\Pi^{H^0}(k^2)$ in 
Eqs. (\ref{eq:dZHkk}, \ref{eq:dZHlk}, \ref{eq:dZAG}) include 
momentum-independent contributions from the 
tadpole shifts~\cite{pokorski,dabelstein} and leading 
higher-order corrections. 
The latter contribution is relevant for the corrections to 
($m_{h^0},m_{H^0},\alpha$). 
For the $A^0$ decays, Eq.~(\ref{eq:dZAG}) already
includes the contribution from the $A^0-Z^0$ mixing in addition to 
the $A^0-G^0$ mixing, 
using the Slavnov-Taylor identity, $\Pi^{H^0}_{43}(m_{A^0}^2) = 
i\,\frac{m_{A^0}^2}{m_{Z^0}} \Pi_{AZ}(m_{A^0}^2)$. 
The explicit forms of the self energies $\Pi^{\ch}(p^2)$, 
$\Pi_{ab}^{H^0}(p^2)$, and 
$\Pi_{AZ}(m_{A^0}^2)$ are shown, for example, in 
Refs.~\cite{pierce,weber_A0sf2}. 

To obtain ultraviolet finite corrections, we further need the counter 
term contribution $F^{(c)}_{ijk}$ from the renormalization of 
the parameters in the tree-level couplings Eq.~(\ref{eq:Fchtree}). 
The chargino mixing matrices ($U$, $V$) 
%and the Higgs mixing angles ($\alpha$, $\beta$) 
are renormalized in the 
on-shell scheme, as described in Refs.~\cite{chmasscorr1,chmasscorr2}. 
In this scheme, extending Ref.~\cite{earlier} for quark and lepton mixings, 
the counter terms for ($U$, $V$) are determined 
such as to cancel the antihermitian parts of the chargino 
wave-function corrections Eq.~(\ref{eq:dZchps}). As a result, 
after including ($\delta V$, $\delta U$) into Eq.~(\ref{eq:dZchps}), 
$\delta{Z}^{+{L,R}}_{i'i}$ are modified as 
$(\delta{Z}^{+{L,R}}_{i'i}+\delta{Z}^{+{L,R}}_{ii'})/2$. 
%Similarly, the on-shell renormalization of the mixing angle $\alpha$ 
%for ($h^0$, $H^0$) results in the symmetrization of Eq.~(\ref{eq:dZHlk}) 
%as $\delta{Z}^{H^0}_{ab} \to (\delta{Z}^{H^0}_{ab}+\delta{Z}^{H^0}_{ba})/2$. 
The counter term of $\beta$ for $A^0$ decays is fixed by the 
condition \cite{pokorski,dabelstein} that the 
renormalized $A^0-Z^0$ mixing self energy $\Pi_{A^0Z^0}(p^2)$ vanishes 
at $p^2=m_{A^0}^2$. Inclusion of this counter term $\delta\beta$ 
cancels the half of $\delta{Z}^{H^0}_{43}$ in Eq.~(\ref{eq:dZAG}).  
As usual, we use the pole mass $m_{A^0}$ and 
on-shell $\tan\beta$ as inputs for the Higgs boson sector. 

Since the zero-momentum contribution $\Pi^{H^0}_{kl}(0)$ to 
the masses and mixing angle of ($h^0$, $H^0$) are often very large, 
we calculate ($m_{h^0}$, $m_{H^0}$) and 
the effective mixing angle $\alpha_{\rm eff}$, 
which is defined to cancel the zero-momentum part of 
$\Pi^{H^0}_{ab}(p^2)$ in Eq.~(\ref{eq:dZHlk}), by FeynHiggs \cite{feynhiggs}, 
which includes the leading higher-order 
corrections, and use these values both for the tree-level and 
corrected widths. After the inclusion of the corresponding counterterm 
$\delta\alpha$, Eq.~(\ref{eq:dZHlk}) is modified as 
\begin{equation}
\label{eq:dZHlk2}
\delta{Z}^{H^0}_{ab} \to \frac{2}{m^2_{H^0_a}-m^2_{H^0_b}}\, 
{\rm Re}\, 
\left[ \Pi^{H^0}_{ab}(m^2_{H^0_b}) - \Pi^{H^0}_{ab}(0) \right] \, , 
\hspace{10mm} a,b=(1,2), \;  a\neq b
\end{equation} 
%(with the UV parameter $\Delta$ set to zero and 
%the scale $Q$ is set to $m_Z$).
with the $\DRbar$ renormalization scale $Q = m_Z$ for $\Pi^{H^0}_{ab}(p^2)$.
%%%%%%%%%%%%%
%If one is replacing $\alpha$ in all the previous formulae
%with the effective one, the self energies $\Pi^{H^0}_{kl}(k^2)$ in the
%wave-function correction and $\delta\alpha$ must be replaced by
%$\Delta\Pi^{H^0}_{kl}(k^2)=\Pi^{H^0}_{kl}(k^2)-\Pi^{H^0}_{kl}(0)$.
%Nevertheless, the form of their sums eqs.~(\ref{eq:dZHkk},\ref{eq:dZHlk}) is
%not affected by the elimination of
%$\Pi^{H^0}_{kl}(0)$.)

Our calculation is performed in the $\xi=1$ gauge. 
Although the on-shell mixing matrices generally depend on the gauge 
parameter \cite{gaugedep,tanbgauge}, our ($U$, $V$) may 
be understood as the ones improved by the 
pinch technique \cite{pinch,espinosa}. 
We ignore here very small differences of the on-shell 
%$\alpha$ and 
$\beta$ between the $\xi=1$ results and 
improved ones by the pinch technique (see Refs.~\cite{espinosa,pinchhiggs} 
for the case of CP-even Higgs bosons). 

For the renormalization of the SU(2) gauge coupling $g$ in 
Eq.~(\ref{eq:Fchtree}), two schemes are used.
In both the W- and Z-pole masses $m_W$ and $m_Z$ are input parameters.
The Weinberg angle is defined by $\cos\theta_W = m_W/m_Z$ \cite{sirlin}, 
and therefore
\begin{equation}
\frac{\d \sin\theta_W}{\sin\theta_W} = \frac{\cos^2\theta_W}{\sin^2\theta_W}
\left( \frac{\d m_Z}{m_Z} - \frac{\d m_W}{m_W}\right)\, .
\end{equation} 
In the $\alpha(m_Z)$ scheme we use as input the \msbar running 
electromagnetic coupling $\alpha(m_Z)$ ($= e^2(m_Z)/(4 \pi)$). 
We have 
\begin{equation}
g = \frac{e(m_Z)}{\sin\theta_W}\, , \quad {\rm and} 
\quad \frac{\d g}{g} = \frac{\d e}{e} - \frac{\d \sin\theta_W}{\sin\theta_W}\, ,
\end{equation}
with $\d e$ given e.~g. in \cite{weber_A0sf,neuprod}, 
$\d m_Z$ and $\d m_W$ in \cite{weber_A0sf2}.\\ 
In the other scheme, called here the $G_F$ scheme,
the Fermi constant $G_F$ for the muon decay is input parameter, 
\begin{equation}
g = \left[ \frac{8G_F m_W^2}{\sqrt{2}} \right]^{1/2}\, ,  
\quad {\rm and} \quad  \frac{\d g}{g} = \d Z_e - \frac{1}{2} \Delta r - \frac{\d \sin\theta_W}{\sin\theta_W}
\end{equation}
$\d Z_e$ is the renormalization constant for the electric charge in the 
Thomson limit \cite{denner}.
The term $\Delta r$ includes the full one-loop MSSM 
correction \cite{deltarSUSY} 
and the leading two-loop QCD corrections \cite{deltar2loopaas}.

%We fix the electroweak gauge boson sector by $m_Z$, $m_W$, and $e$.
%One gets from the relations $g=e/s_W$, $g'=e/c_W$, and 
%($c_W\equiv \cos\theta_W$, $s_W \equiv \sin\theta_W$)~\cite{sirlin,denner}
% \begin{equation}
%  \frac{\delta g}{g}=  \frac{\delta e}{e} +\frac{c_W^2}{2\,s_W^2}
% \left(\frac{\delta m_W^2}{m_W^2}-\frac{\delta m_Z^2}{m_Z^2}\right)\, ,
% \label{eq:dggdgpgp}
% \end{equation}
%The formulae for $\delta m_W$ and $\delta m_Z$ can be also found in
%\cite{chmasscorr} and for $\d e/e$ in the Appendix~\ref{sec:appB}.

The corrected widths are
\begin{eqnarray}
\Gamma^{\rm corr} 
 &=& \Gamma^{\rm tree} +  \frac{g^2}{16 \pi\, m^{3}_{H_{k}^0} }\,
\kappa(m_{H_{k}^0}^2,m^{2}_i,m^{2}_j) \,
\left[ 
\left( m^{2}_{H_{k}^0} - m^{2}_i - m^{2}_j \right) 2 {\rm Re}
(F_{ijk} \Delta F_{ijk} + F_{jik} \Delta F_{jik} ) 
\right. 
\nonumber \\ 
&& \left. 
 - 4 \eta_k m_i m_j {\rm Re}
( F_{ijk} \Delta F_{jik} + F_{jik} \Delta F_{ijk} )  \right] 
\nonumber\\ 
&& + \Gamma(H_{k}^0 \to \ch^+_i\, \ch^-_j\, \gamma ) \, . 
\label{eq:gammacorr}
\end{eqnarray}
The process $H^0_k\to\ch^+_i\ch^-_j\gamma$ with real photon emission 
is included to cancel the infrared divergence by virtual photon loops.

One has to be careful in using the on-shell mixing matrices ($U$, $V$) 
and masses $m_i\,(i=1,2)$ in the numerical analysis. 
When the gauge and Higgs boson sectors are fixed, 
the chargino sector is fixed by two independent parameters. 
Here we follow the method proposed in Refs.~\cite{chmasscorr1,chmasscorr2}: 
We fix the chargino sector by taking
$M \equiv X_{11}$ and 
$\mu \equiv X_{22}$,
where the on-shell mass matrix $X$ is defined to give the on-shell 
masses $m_i$ and on-shell mixing matrices ($U$, $V$) by diagonalization. 
Note that, for given values of the on-shell $M$ and $\mu$, 
the one-loop corrected on-shell masses $m_i$ and mixing matrices 
($U$, $V$) are shifted~\cite{chmasscorr1,chmasscorr2} from 
the values obtained by the tree-level 
mass matrix $X^{\rm tree}$ composed by the input parameters, the
on-shell $M, \mu, \tan\beta$, and the pole mass $m_W$. 
This is due to the shift of the off-diagonal elements of 
$X$ from their tree-level values and related to 
the deviation of the gaugino couplings from 
the corresponding gauge couplings by SUSY-breaking 
loop corrections~\cite{superoblique}. 
These shifts of $m_i$ and ($U$, $V$), in addition to the 
``conventional'' corrections shown in Eq.~(\ref{eq:gammacorr}), 
have to be taken into account 
for a proper treatment of the loop corrections. 
(A slightly different scheme for the chargino sector 
was proposed in Ref.~\cite{chmasscorr3}. 
Apart from the different definition of the renormalized $M$ and $\mu$, 
their method is equivalent to ours.)

The full one-loop corrections were calculated using the packages 
FeynArts, FormCalc, and LoopTools \cite{FeynArts}. For the contributions 
of the quarks, leptons, and their superpartners, we also checked 
the consistency with Ref.~\cite{Zhang}, both analytically 
and numerically.

\section{Numerical results}

We present numerical results for the tree-level and one-loop 
widths of the decays $A^0\to\ch^+_1\ch^-_1$,
$H^0\to\ch^+_1\ch^-_1$, and $\chi_2^+ \to \chi_1^+ h^0$. 
The SUSY parameter set SPS1a of the 
Snowmass Points and Slopes in Ref.~\cite{SPS1a} is 
chosen as reference point; 
For the trilinear breaking terms $A_t$, $A_b$ and $A_\tau$ we
use the $\drbar$ running values given at the scale of the mass of 
the decaying particle, $A_t = -487$~GeV, $A_b = -766$~GeV, $A_\tau = -250$~GeV.
All other parameters are taken on-shell,
$M=197.6$~GeV, $M'=98$~GeV, 
$\mu=353.1$~GeV, $\tan\beta=10$, and $m_{A^0} = 393.6$~GeV. 
The soft breaking sfermion mass parameters,
for the first and second generation are 
$M_{\tilde Q_{1,2}} = 558.9$~GeV, 
$M_{\tilde U_{1,2}} = 540.5$~GeV,
$M_{\tilde D_{1,2}} = 538.5$~GeV,
$M_{\tilde L_{1,2}} = 197.9$~GeV,
$M_{\tilde E_{1,2}} = 137.8$~GeV,
and for the the third one, 
$M_{\tilde Q_{3}} = 512.2$~GeV, 
$M_{\tilde U_{3}} = 432.8$~GeV,
$M_{\tilde D_{3}} = 536.5$~GeV,
$M_{\tilde L_{3}} = 196.4$~GeV,
$M_{\tilde E_{3}} = 134.8$~GeV. 
In all figures, 
these values are used, if not specified otherwise.

For the standard model parameters, we take $\alpha(m_Z) = 1/127.922$, 
$m_Z=91.1876$~GeV,
$m_W=80.423$~GeV, the on-shell parameters $m_t=174.3$~GeV, 
and $m_\tau = 1.777$~GeV. 
For the bottom mass,
our input is the $\MSbar$ value $m_b(m_b) = 4.2$~GeV.
For the values of the Yukawa couplings of 
the third generation quarks ($h_t$, $h_b$), we take the
running ones at the scale of the decaying particle mass.\\
In the $G_F$ scheme for the renormalization of $g$, we use 
$G_F=1.16639\times 10^{-5}$~GeV$^{-2}$ instead of $\alpha(m_Z)$. 

We compare three cases: the ``naive'' tree-level width $\Gnaive$, 
the tree-level width already including the loop corrections to the 
chargino mass matrix $\Gtree$, 
and the full one-loop width $\Gcorr$.

\begin{figure}[h!]
 \begin{center}
\hspace{2mm}
\mbox{\resizebox{75mm}{!}{\includegraphics{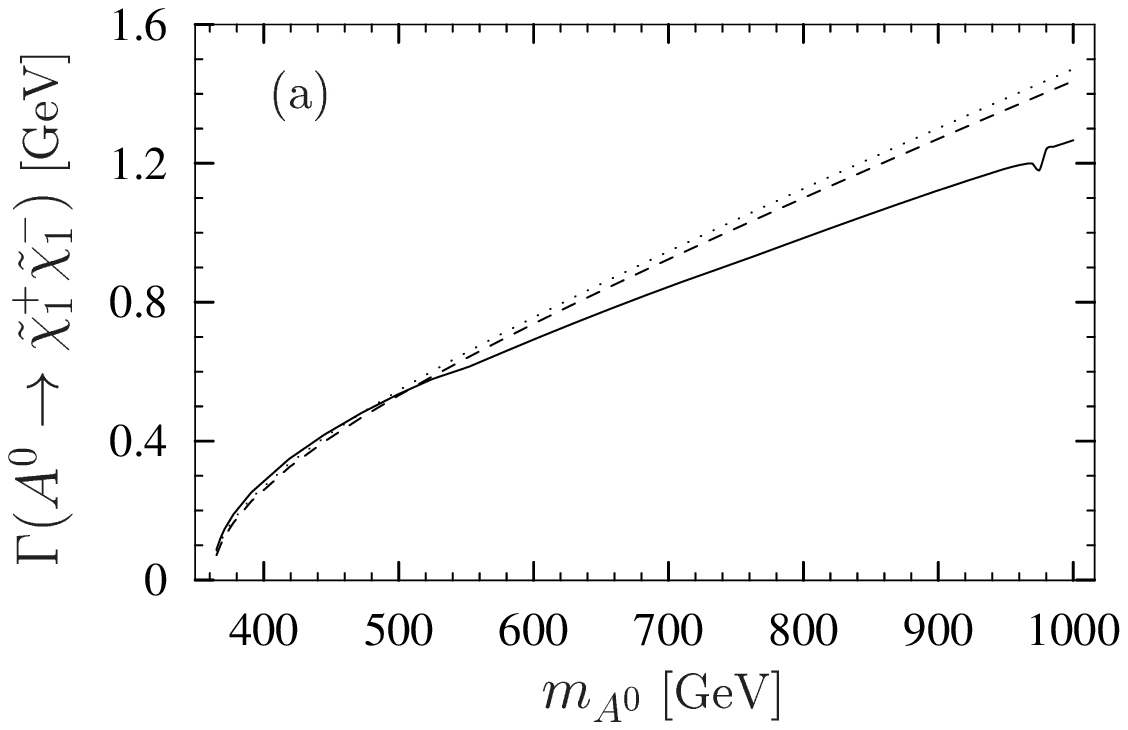}}}
\hfill
\mbox{\resizebox{75mm}{!}{\includegraphics{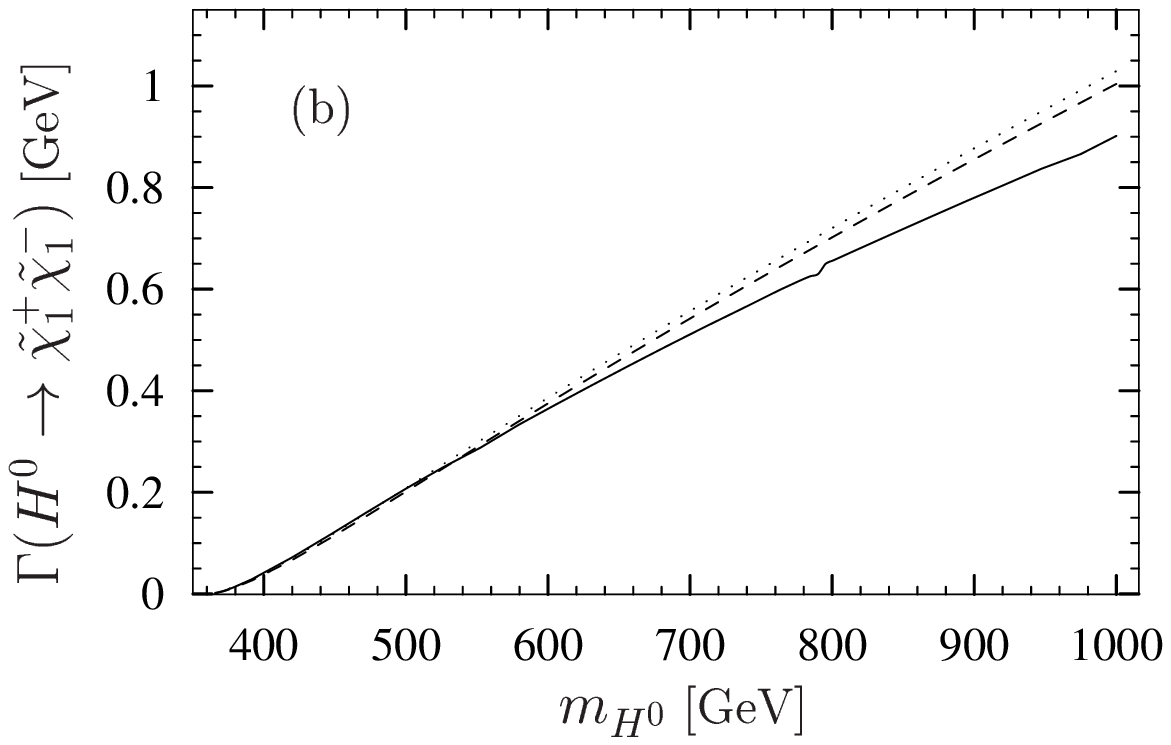}}}\\
\mbox{\resizebox{83mm}{!}{\includegraphics{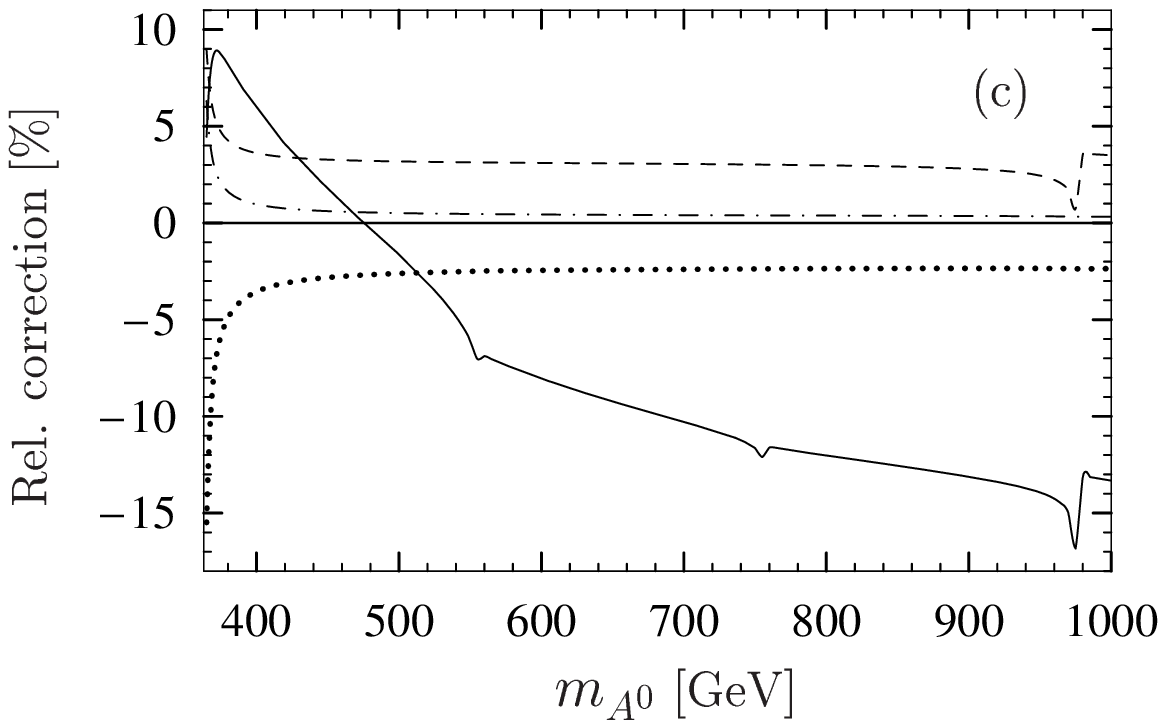}}}
\hspace*{-5mm}
 \caption[fig2]
 {Naive tree-level (dotted), tree-level (dashed) and 
  one-loop corrected (solid) widths of 
  the decays $A^0\to\ch^+_1+\ch^-_1$ as functions of $m_{A^0}$ (a),  
  and $H^0\to\ch^+_1+\ch^-_1$ as functions of $m_{H^0}$ (b), 
  in the $\alpha(m_Z)$ schemes for the renormalization 
  of the SU(2) gauge coupling $g$. The individual
 loop contributions to (a) are shown in (c), for explanation 
 see the text.}
\label{fig:mHdependence}
 \end{center}
 \vspace{-5mm}
 \end{figure}

In Fig.~\ref{fig:mHdependence} we show
the tree-level and corrected widths in (a) of $A^0\to\ch^+_1\ch^-_1$ 
as functions of $m_{A^0}$, and
in (b) of $H^0\to\ch^+_1\ch^-_1$  
as functions of $m_{H^0}$. 
The tree-level branching ratios of these decays at $m_{A^0}=393.6$~GeV 
(where $m_{H^0}=394.1$~GeV) are, using HDECAY program \cite{HDECAY}, 
${\rm Br}(A^0\to\ch^+_1\ch^-_1)=21$~\% and 
${\rm Br}(H^0\to\ch^+_1\ch^-_1)=4$~\%, which are not negligible. 
We see that the full one-loop corrections amount up to $\sim -12$\%.
In Fig.~\ref{fig:mHdependence}~(c) the individual contributions to
Fig.~\ref{fig:mHdependence}~(a) relative to the naive tree-level width
are exhibited. The dash-dotted line show the (s)fermion loop 
contribution (loops with quarks, leptons, and their superpartners) 
through the correction to the chargino mass matrix, while 
the dotted line shows the full correction to the mass matrix. 
The solid (dashed) line shows the total correction 
$\Gcorr/\Gnaive-1$ including full ((s)fermion) one-loop contributions. 
This figure shows that the (s)fermion loop corrections and 
other corrections are of comparable order, both for the chargino 
mass matrix and for the conventional corrections (\ref{eq:gammacorr}). 

\begin{figure}[h!]
 \begin{center}
\mbox{\resizebox{75mm}{!}{\includegraphics{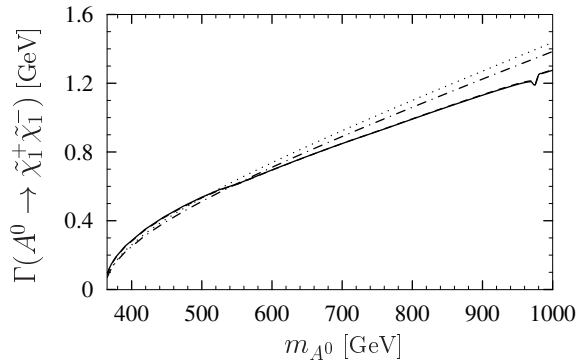}}}
 \caption[fig3]
 {Comparison of the results using the $\alpha(m_Z)$ scheme 
or the $G_F$ scheme for the decay widths of $A^0\to\ch^+_1\ch^-_1$. 
The dotted and the solid (dash-dotted and dashed) lines denote
the tree-level and one-loop corrected line in the $\alpha(m_Z)$ ($G_F$) scheme.
}
 \label{fig:g_schemes}
 \end{center}
 \vspace{-5mm}
 \end{figure}

A comparison of two renormalization schemes for fixing $g$, 
the $\alpha(m_Z)$ scheme 
and the $G_F$ scheme, is shown in Fig.~\ref{fig:g_schemes} for the decay 
$A^0\to\ch^+_1\ch^-_1$ as functions of $m_{A^0}$. 
The difference between these two schemes is below 1\%, 
scaling with the one-loop correction part, and 
mainly a higher order effect.

\begin{figure}[h!]
 \begin{center}
\hspace{2mm}
\mbox{\resizebox{75mm}{!}{\includegraphics{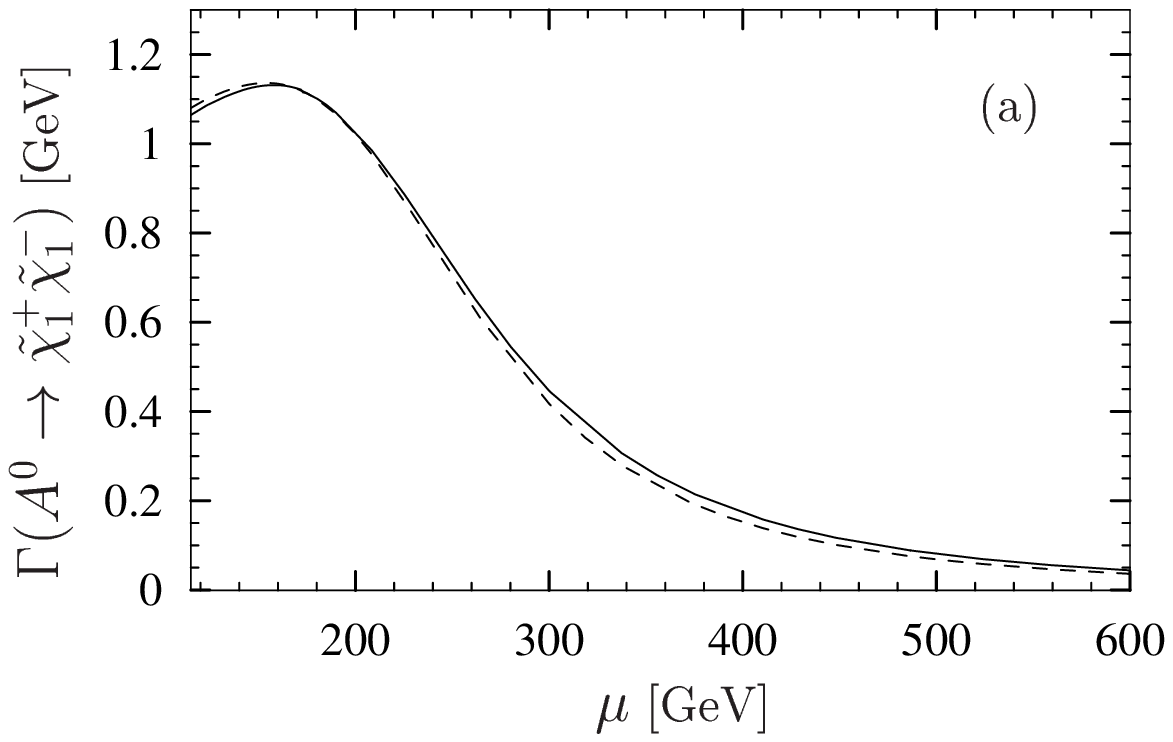}}}
\hfill
\mbox{\resizebox{75mm}{!}{\includegraphics{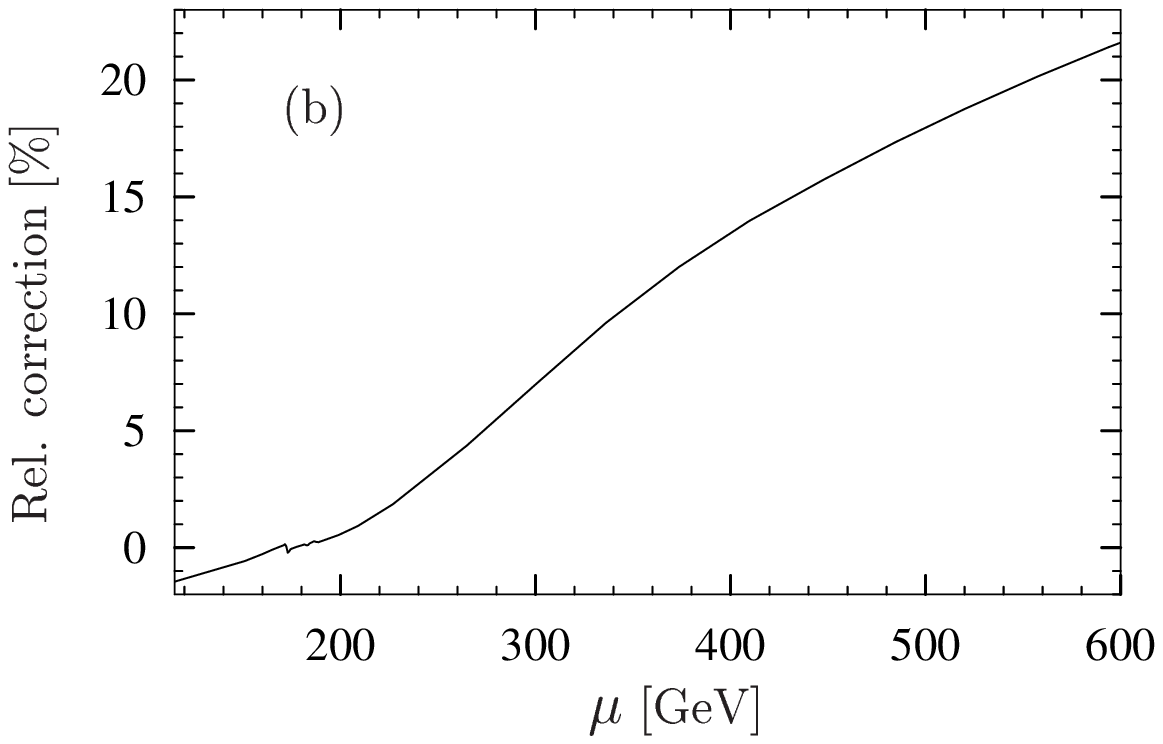}}}
\hspace*{-5mm}
 \caption[fig4]
 {Tree-level (dotted) and one-loop corrected (solid) widths of 
the decays 
$A^0\to\ch^+_1+\ch^-_1$~(a) and (b) the correction of this process 
relative to the tree-level width as a function of $\mu$}
 \vspace{-5mm}
 \label{fig:mudependence}
 \end{center}
\end{figure}

Since the Higgs boson couplings to charginos are very sensitive to 
the gaugino-higgsino mixing, it is interesting to study the dependence of 
the decay widths on the gaugino and higgsino components of $\ch^{\pm}_1$. 
Fig.~\ref{fig:mudependence} shows the 
tree-level and one-loop corrected widths of $A^0\to\ch^+_1+\ch^-_1$ 
as functions of $\mu$ for fixed $M$. 
One can see that in the region where the light chargino $\ch^+_1$ becomes 
a pure wino the width gets very small. 
The correction grows from $\sim -1$\% for $\mu\sim 120$~GeV 
to 20\% for $\mu \sim 600$~GeV. 
The $\mu$ dependence of the decay width $H^0\to\ch^+_1+\ch^-_1$ is 
not shown because its behavior is similar to
that shown in Fig.~\ref{fig:mudependence}. 

\begin{figure}[h!]
 \begin{center}
\hspace{2mm}
\mbox{\resizebox{75mm}{!}{\includegraphics{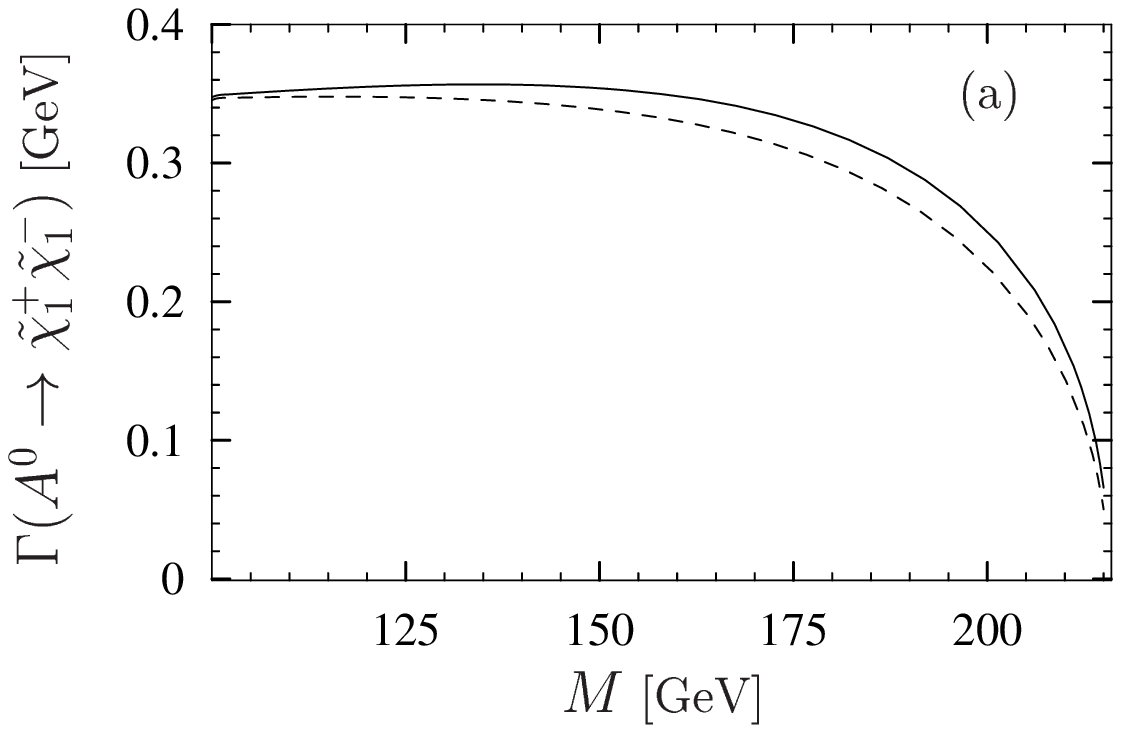}}}
\hfill
\mbox{\resizebox{75mm}{!}{\includegraphics{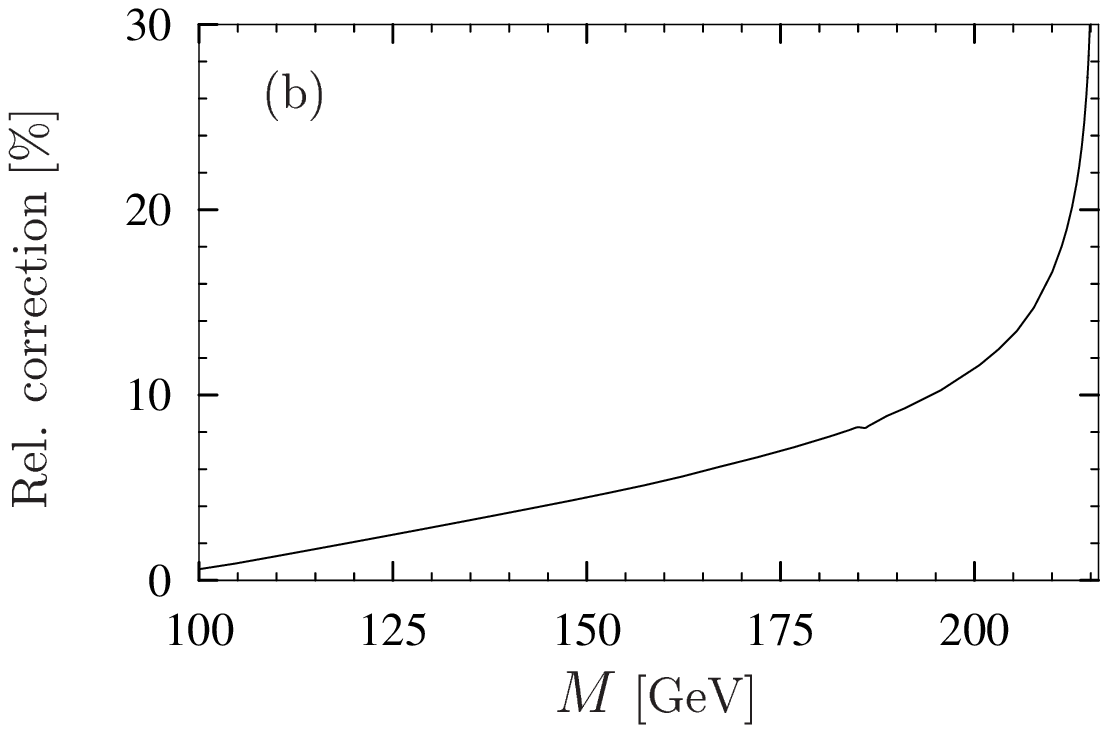}}}
\hspace*{-5mm}
\caption[fig5]
{Tree-level (dotted) and one-loop corrected (solid) widths of the
decay $A^0\to\ch^+_1+\ch^-_1$~(a) and (b) the corrections of 
this process relative to
the tree-level widths as a function of $M$.}
\label{fig:Mdependence}
\end{center}
\vspace{-5mm}
\end{figure}

Fig.~\ref{fig:Mdependence} shows the 
tree-level and one-loop corrected widths of $A^0\to\ch^+_1+\ch^-_1$ 
as functions of $M$ for fixed $\mu$.
In the whole range of this figure $\ch^+_1$ is gaugino-like.
In (a), for increasing $M$ the decay widths decreases due to phase space.
The correction, see (b), gets up to 30\% near the threshold. Again, 
the $H^0\to\ch^+_1+\ch^-_1$ is not shown
because of a similar behavior. 

\begin{figure}[h!]
 \begin{center}
\hspace*{-3mm}
\mbox{\resizebox{78mm}{!}{\includegraphics{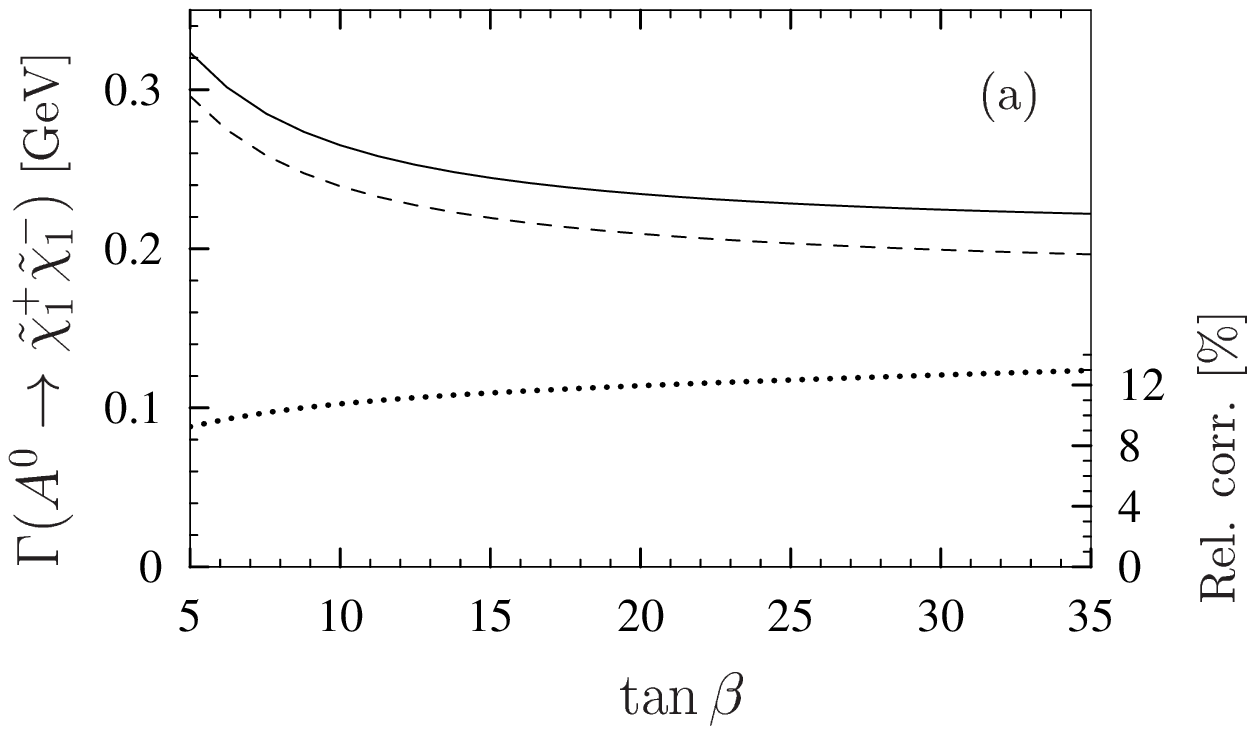}}}
%\hspace{5mm}
\hfill
\mbox{\resizebox{78mm}{!}{\includegraphics{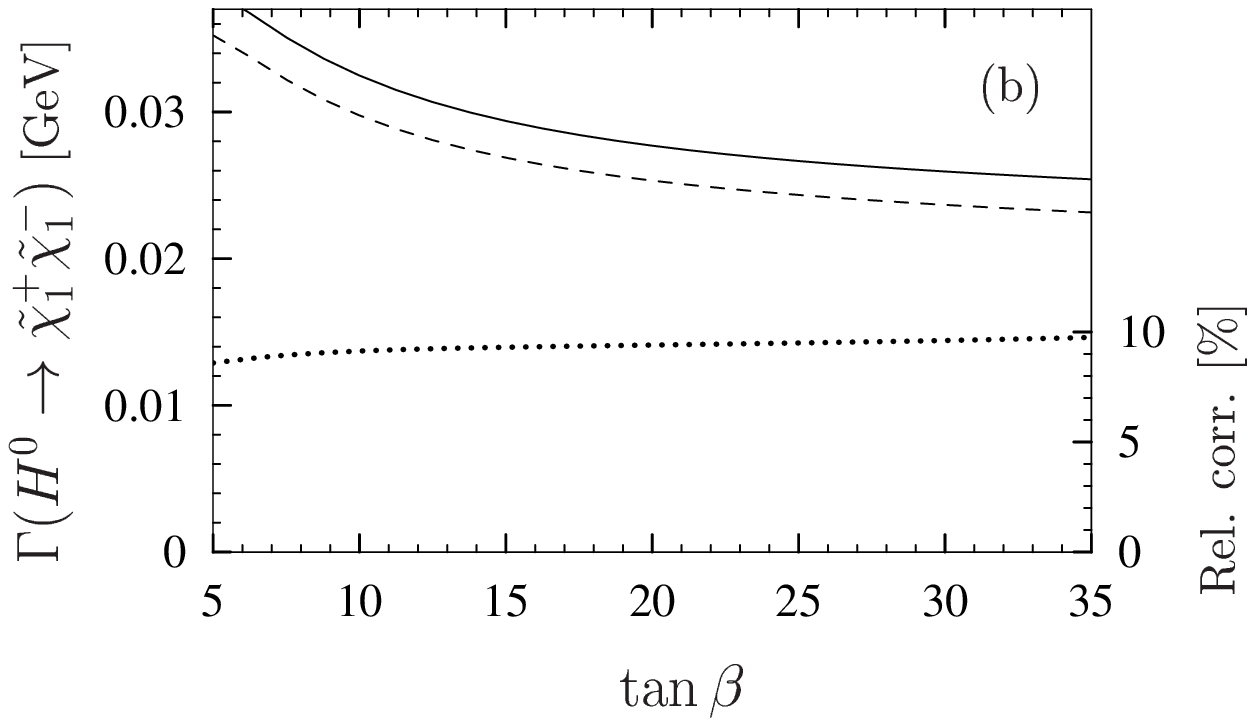}}}
\hspace*{-3mm}
\caption[fig6]
{Tree-level (dashed), one-loop corrected (solid) width and 
the correction (dotted) 
relative to the the tree-level width for the decays 
$A^0\to\ch^+_1+\ch^-_1$~(a) and $H^0\to\ch^+_1+\ch^-_1$~(b) 
as a function of $\tan\beta$.}
\label{fig:tanbdependence}
 \end{center}
 \vspace{-5mm}
 \end{figure}

Fig.~\ref{fig:tanbdependence} shows the decay widths for  
$A^0\to\ch^+_1+\ch^-_1$ in (a) and  $H^0\to\ch^+_1+\ch^-_1$ in (b) 
as functions of $\tan\beta$. The correction 
is in the range of $\sim 10$\% and the dependence on 
$\tan\beta$ is small. We examined the difference of the renormalization scheme 
taking the $\DRbar$ value for $\tan\beta$ at the scale $Q = 454.7$~GeV as input
parameter instead of the on-shell $\tan\beta$. For these processes
the difference is small, e.g. in the Fig.~\ref{fig:tanbdependence}~(a) it is
about 0.5\% for low and 0.2\% for large $\tan\beta$, respectively.

\begin{figure}[h!]
\begin{center}
\hspace*{2mm}
\mbox{\resizebox{73mm}{!}{\includegraphics{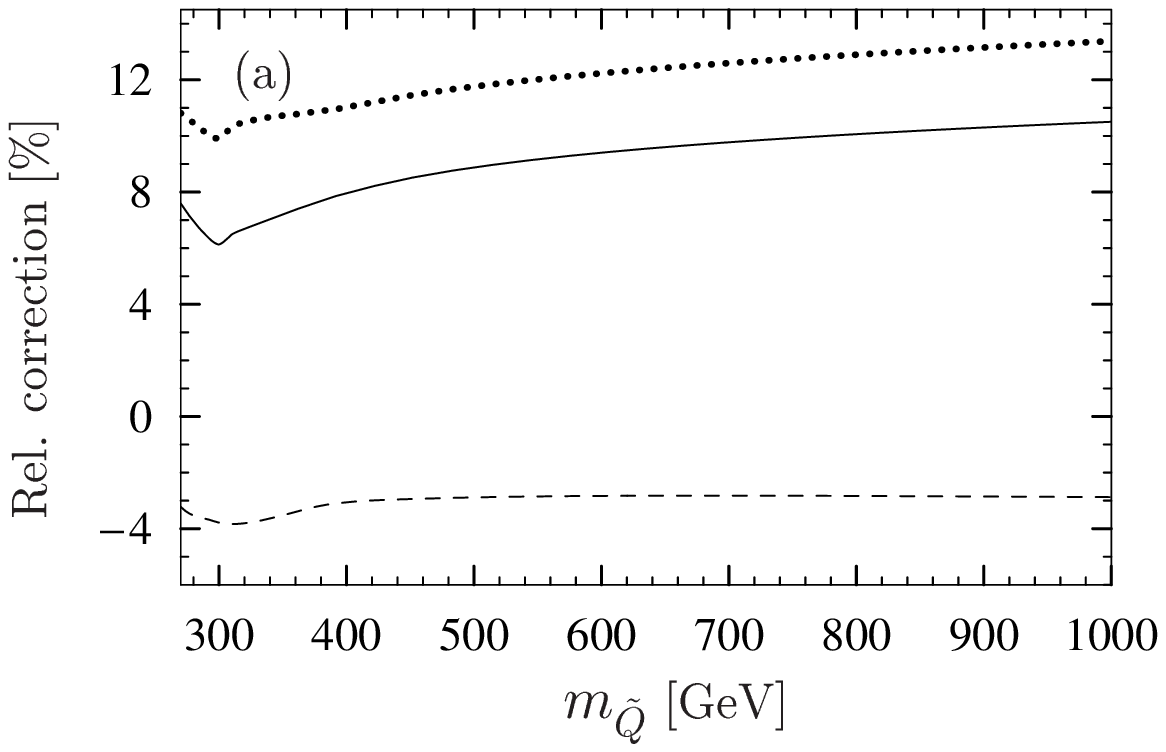}}}
%\hspace{5mm}
\hfill
\mbox{\resizebox{73mm}{!}{\includegraphics{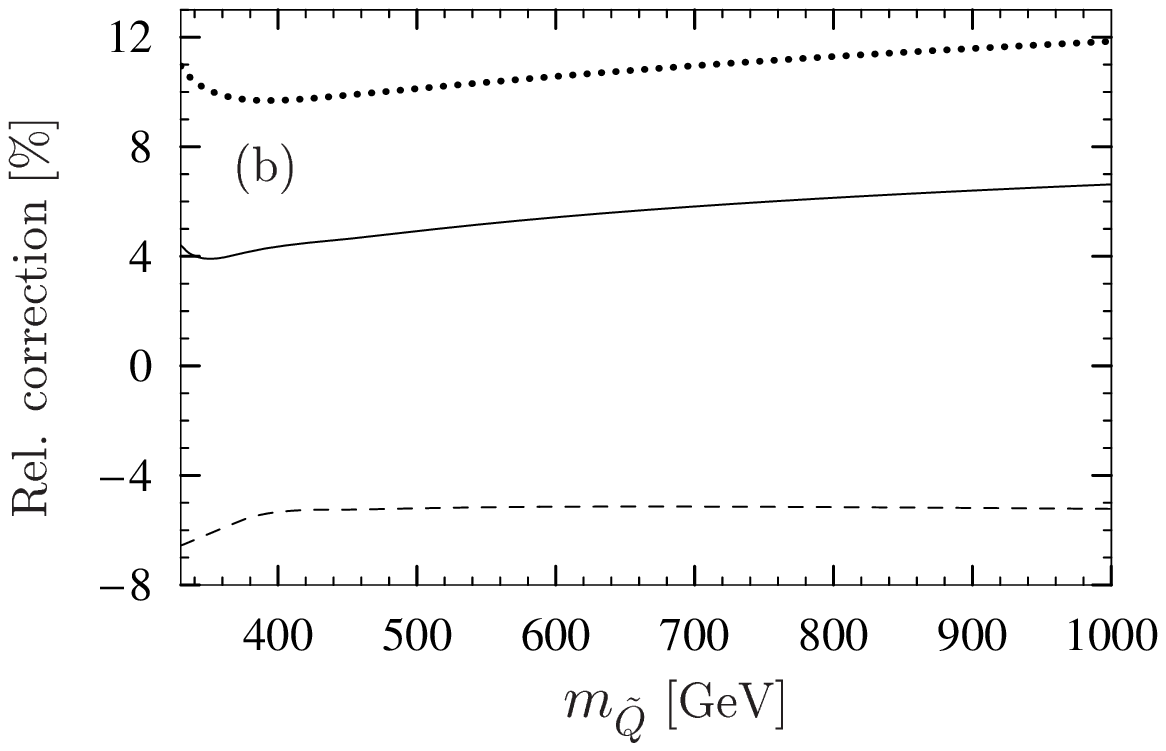}}}
\hspace*{-1mm}
\caption[fig7]
{Correction of the full one-loop corrected (solid),  
the tree-level (dashed), and the conventional one-loop corrected 
width (dotted) for the decays 
$A^0\to\ch^+_1+\ch^-_1$~(a) and $H^0\to\ch^+_1+\ch^-_1$~(b)  
relative to the naive tree-level width as a function of $m_{\tilde Q}$. 
(Note that the tree-level already includes the correction due to the 
chargino mass matrix renormalization.)}
\label{fig:msQdependence}
\end{center}
\vspace{-5mm}
\end{figure}

Fig.~\ref{fig:msQdependence} shows the corrections to the decay widths for  
$A^0\to\ch^+_1+\ch^-_1$ in (a) and  $H^0\to\ch^+_1+\ch^-_1$ in (b)
relative to the naive tree-level width as functions 
of $ m_{\tilde Q}$.  The SUSY breaking mass terms for 
all sfermions ($M_{\tilde Q_i},M_{\tilde U_i},M_{\tilde D_i}, 
M_{\tilde L_i},M_{\tilde E _i}$) ($i=1,2,3$) are taken to be equal 
to $ m_{\tilde Q}$, 
while the other parameters are unchanged.
The relative corrections $\Gtree/\Gnaive-1$ (dashed lines), stemming from 
the shift of the chargino mass matrix by the renormalization, 
are negative. The remaining conventional corrections 
shown in Eq.~(\ref{eq:gammacorr}) (dotted lines) are positive. 
The total correction $\Gcorr-\Gnaive$ (solid lines) is positive and 
in the range of $6 - 11$\% in (a) and $4 - 7$\% in (b). 
The corrections become quite insensitive to $m_{\tilde Q}$ 
for large $m_{\tilde Q}$. 
The total correction consists of the $m_{\tilde Q}$ dependent 
(s)fermion contribution and the remaining contribution, 
the latter of which is $\sim 7.8$\% for (a) and 
$\sim 9.6$\% for (b). 
Again, these two types of loop corrections are of comparable order. 

\begin{figure}[h!]
\begin{center}
\hspace*{2mm}
\mbox{\resizebox{73mm}{!}{\includegraphics{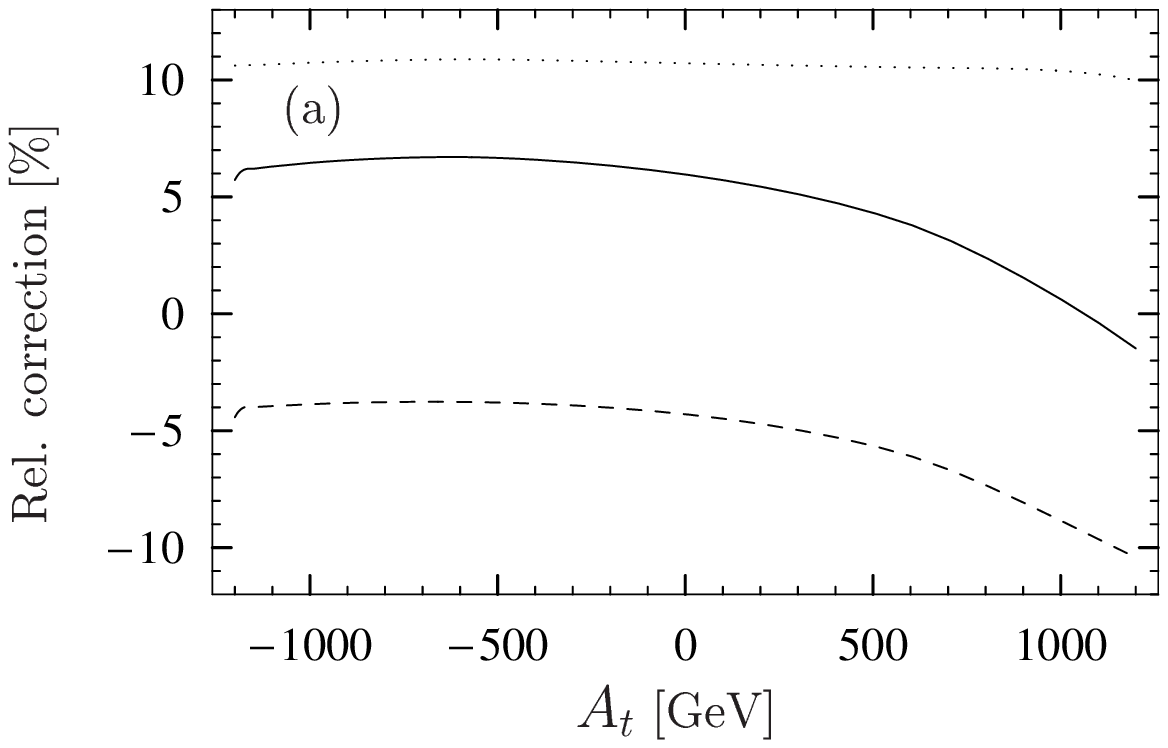}}}
%\hspace{5mm}
\hfill
\mbox{\resizebox{73mm}{!}{\includegraphics{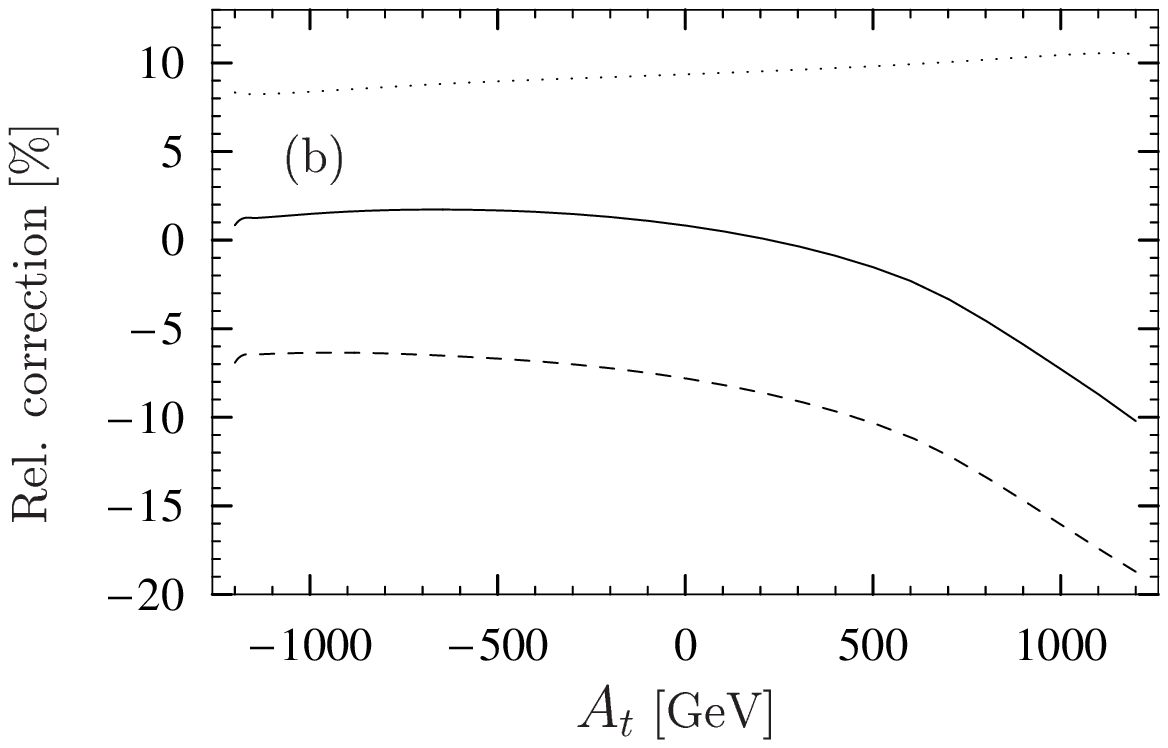}}}
\hspace*{-1mm}
\caption[fig7]
{Relative corrections for the decays 
$A^0\to\ch^+_1+\ch^-_1$~(a) and $H^0\to\ch^+_1+\ch^-_1$~(b) as a function 
of $A_t$. The dashed lines denote $\Gtree/\Gnaive - 1$, the
solid lines denote $\Gcorr/\Gnaive - 1$ and the dotted lines 
$\Gcorr/\Gtree - 1$.}
\label{fig:Adependence}
\end{center}
\vspace{-5mm}
\end{figure}

Fig.~\ref{fig:Adependence} shows the corrections to the decay widths for  
$A^0\to\ch^+_1+\ch^-_1$ in (a) and  $H^0\to\ch^+_1+\ch^-_1$ in (b)
as a function of $A_t = A_b = A_\tau$, 
with the other parameters unchanged.
The dashed lines denote $\Gtree/\Gnaive - 1$. They show the
effect due to the chargino mass matrix renormalization.
The solid lines show the total correction in terms of the naive tree level width,
$\Gcorr/\Gnaive - 1$. The dotted lines stand for $\Gcorr/\Gtree - 1$. 
This is the total correction in terms
of the tree-level result, where the chargino mass matrix renormalization effect
is already included. One sees that $\Gtree/\Gnaive - 1$ and  
$\Gcorr/\Gnaive - 1$ are much stronger dependent on $A_t$ compared to 
$\Gcorr/\Gtree - 1$. 
This shows that the $A_t$ dependence of the corrected widths comes 
mainly from the shifts of the masses and mixing matrices of the 
charginos. 

\begin{figure}[h!]
 \begin{center}
 \hspace{2mm}
\mbox{\resizebox{75mm}{!}{\includegraphics{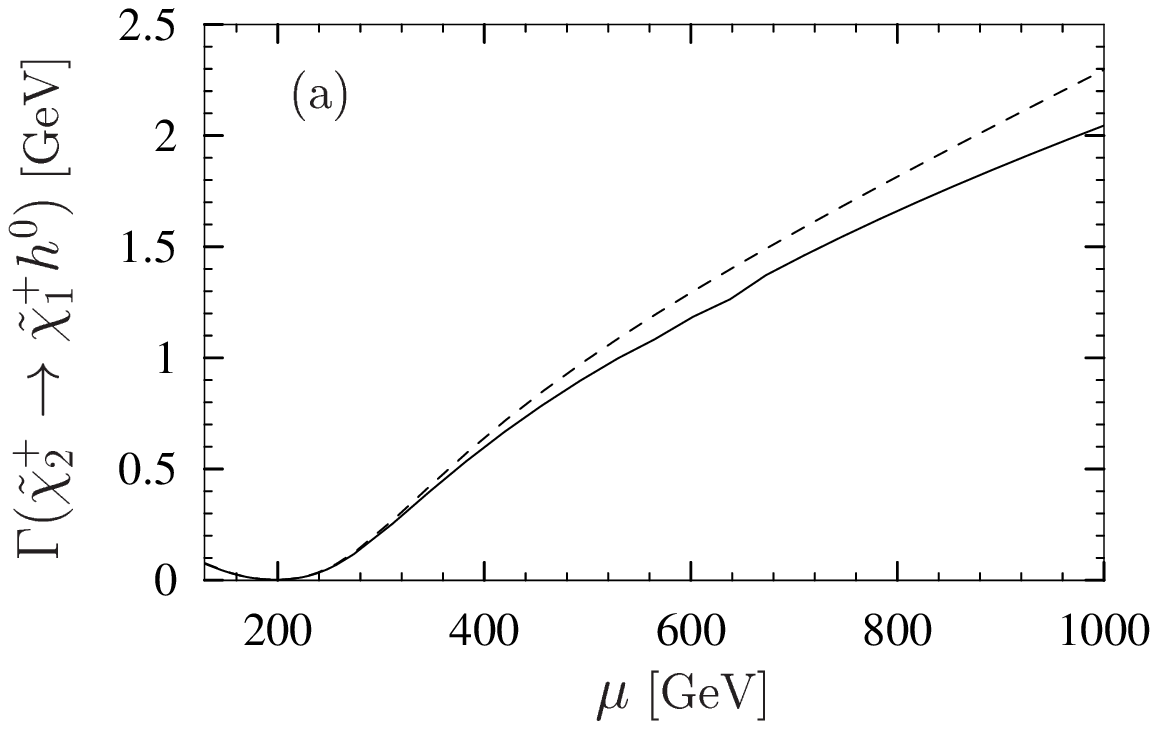}}}
\hfill
\mbox{\resizebox{75mm}{!}{\includegraphics{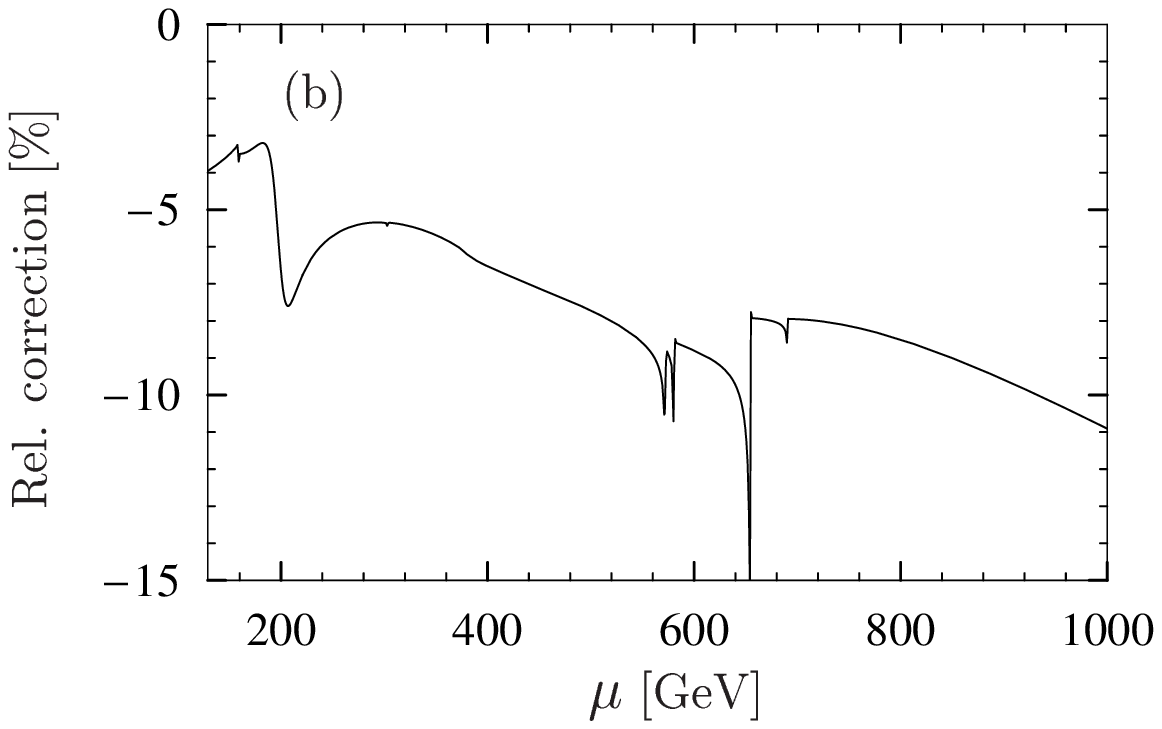}}}
\hspace*{-5mm}
 \caption[fig8]
 {The tree-level and one-loop corrected widths of the
 decay $\ch^+_2\to\ch^+_1h^0$ for varying $\mu$. 
 The dotted and solid lines correspond to the tree-level and loop-corrected 
 widths, respectively.}
 \label{fig:ChaDecay}
 \end{center}
 \vspace{-5mm}
 \end{figure}

Finally, Fig.~\ref{fig:ChaDecay} shows the width of the 
crossed channel decay $\ch^+_2\to\ch^+_1h^0$, as a function of $\mu$. 
The total correction is in the  range of $-5$\% to $-10$\%.
In Fig ~\ref{fig:ChaDecay}~(b) a few pseudo thresholds are seen 
due to opening decay channels into loop particles, such as 
$\ch^+_2\to t \sb_1^*$ at $\mu\sim 650$~GeV. 

\section{Conclusions }
\label{sec:concl}

We have calculated the full one-loop corrections to
the decays 
$(H^0, A^0) \to \ch^+_i  + \ch^-_j$\\ $(i,j=1,2)$. 
All parameters in the chargino mass matrix $X$ and mixing matrices 
($U$, $V$) are renormalized in the on-shell scheme. 
The importance of the corrections to these matrices, 
in addition to the conventional 
corrections (vertex and wave-function corrections with counter
terms), was emphasized. 
We have studied the dependence of the corrections 
on the SUSY parameters. The corrections to
the widths of the decays $(H^0, A^0) \to \ch^+_1  + \ch^-_1$ 
are of the order of $10$\%, but can be larger near the thresholds. 
The corrections from quarks, leptons, and their superpartners were 
shown to be of similar order of magnitude as the other loop corrections. 
We also showed that the correction to 
the decay $\ch^+_2 \to \ch^+_1 h^0$ can be to $\sim  -10$\%.

\section*{Acknowledgements}
We thank Martin Kincel for collaboration in the early stage of this work.
We also thank Ren-You Zhang for communication about the consistency of 
our result to that in Ref.~\cite{Zhang}. 
This work was supported by EU under the HPRN-CT-2000-00149 network 
programme and the ``Fonds
zur F\"orderung der wissenschaftlichen Forschung'' of Austria,
project no.~P13139-PHY. 
Y.~Y. was also supported by the Grant-in-aid for 
Scientific Research from the Ministry of Education, Culture, 
Sports, Science, and Technology of Japan, No.~14740144.

\newpage

\end{document}